\documentclass[prd,showpacs,preprintnumbers,amsmath,amssymb]{revtex4}

\begin{document}
\title{Discrete Spectrum of Inflationary Fluctuations}
\author{Craig J. Hogan}
\address{Astronomy and Physics Departments, 
University of Washington,
Seattle, Washington 98195-1580}
\begin{abstract}
It is conjectured that inflation, taking account  of quantum gravity, leads to a  discrete 
spectrum of cosmological perturbations, instead of the continuous Gaussian spectrum predicted by standard field theory in an unquantized background. 
Heuristic models of discrete spectra  are discussed,    based on an inflaton mode with self gravity, a lattice of amplitude states, an entangled ensemble of modes, and the holographic or covariant entropy bound.    Estimates are given for the discreteness observable in
cosmic background   anisotropy, galaxy clustering, and   gravitational wave backgrounds.
\end{abstract}
\maketitle

\section{Introduction}

Holographic  entropy
bounds derived from generalized black hole thermodynamics and
statistical
physics\cite{'tHooft:1999bw,'tHooft:1985re,Susskind:1993if,stephens,thooft93,susskind95,Bigatti:1999dp,bousso,bousso02,bekenstein01,bekenstein02}
can  be interpreted to  signify fundamental, universal properties of gravitating quantum systems. In this view,
a deeper structure  underlying gravity and quantum mechanics guarantees  that any physical system is described by a fundamental Hilbert space of
finite dimension, where the maximum entropy (the logarithm of this dimension) is given by the area (in Planck units) of an appropriately defined  bounding surface. 
If so, quantum gravity requires a radical departure from the infinite-dimensional Hilbert space of
 quantum field theory.

Yet field theory is the tool used to describe  inflationary 
perturbations\cite{liddlelyth}---  our  most direct observational contact with quantum fields under conditions extreme enough for single quanta to have   significant  self-gravity.
Maps  of  cosmic background anisotropy  show metric perturbations imprinted with the detailed spatial structure of  quantum
mechanical wavefunctions of  field quanta in the  early universe. This paper analyzes an     effect  implied by holography but    
 not apparent in the field-theory analysis: a discrete quantum spectrum that survives  in classical
observables, such as the  amplitudes and spatial structures of relic large-scale metric perturbations. An earlier paper\cite{Hogan:2002xs} estimated  
discreteness in the amplitude spectrum; the analysis here suggests that observable holographic discreteness may also appear in spatial structure.

Any quantum mechanical system\cite{preskill} is  a ray in a Hilbert space $\cal H$. The Hilbert space is spanned by an orthonormal set of $N_{\cal H}$ basis vectors. Any
observation consists of the projection of the ray onto a complete, orthonormal basis formed by the eigenstates of some self-adjoint operator (an ``observable''), and the
only possible results of any observation are its  eigenvalues in this spectral representation. That is, any observation has only  $N_{\cal H}$ possible outcomes, each of which leaves the
system in an eigenstate of that observable.   A classical state of definite energy, such as a metric perturbation from inflationary fluctuations, corresponds to  one of the eigenstates of
the Hamiltonian. The dimension
$N_{\cal H}$ of the Hilbert space determines  the maximum possible number of  classical outcomes, and also the maximum entropy $S=\log N_{\cal H}$. 
A  bound  on   $S$  implies a finite number of eigenstates and  a discrete spectrum of possible classical outcomes.
 The  particular pattern of eigenvalues of a quantum system (for example, the spectrum of lines of atomic transitions\cite{atominbox}) conveys rich information
about its dynamical elements.  
Similarly, it may be possible to resolve  spectral features of
inflationary perturbations, revealing details of the fundamental dynamical elements of quantum gravity.

The observability of this discreteness depends on the number
of eigenstates spanning the space of the vacuum fluctuations during inflation. This paper estimates this number by analyzing the behavior of a series of simplified  quantum mechanical
models of
    field modes during inflation, and by application of the covariant entropy bound to slow-roll inflation.
The simple quantum systems
considered here are not  serious models of quantum gravity, but   are chosen because they display discrete spectra due to   interactions resembling gravity, they can be
related  directly to inflationary fields, and at the same time they provide a bridge between field behavior and  some of the new physics  of the powerful holographic entropy bound.

Rather than incorporating the fully developed theory of quantum fields during
inflation\cite{Starobinsky:1979ty,Hawking:1982cz,Guth:1982ec,Bardeen:1983qw,Starobinsky:1982ee,Halliwell:1985eu,Grishchuk:1993ds,Grishchuk:1989ss,Grishchuk:1990bj,Albrecht:1994kf,Lesgourgues:1997jc,Polarski:1996jg,kiefer}, the approach here aims 
to gain some insight into   how the self-gravity of quantum fields leads to a finite (rather than infinite) Hilbert space, and how the quantum states  translate  to the spectra of final classical
metric perturbations. There is no reason to expect significant changes in field-theoretic
  predictions for the mean power spectrum of the perturbations\cite{Mukhanov:1992me,lythriotto},   perturbative estimates of stringy and transplanckian influences on
perturbation spectra
\cite{brandenberger01,martin01,starobinsky01,niemeyer,easther,hui,Maldacena:2002vr}, or  interpretation of thermodynamic entropy
flow during inflation\cite{Frolov:2002va,Albrecht:2002xs}.  The continuous formalism of field theory still presumably applies as a good approximation  for computing  the final classical  probabilities for spectral averages such as the power spectrum. However, standard field theory does not predict discrete  perturbation spectra.

 The analysis below is based mostly on the  quantum mechanics of   comoving modes of a massless free field.
A single spatial mode is equivalent to a simple one-dimensional harmonic oscillator with a frequency that decreases as the physical wavelength of the mode increases with the
cosmic expansion.  This system already displays the main effect leading to inflationary perturbations,  converting vacuum-eigenstate fluctuations into  quasi-classical
amplitude eigenstates. The infinite $N_{\cal H}$   corresponds to an infinite number of number eigenstates at early times and a continuous, nearly
Gaussian distribution of measurable amplitude values at late times--- the standard prediction  of inflation.  Adding terms to this Hamiltonian representing the  self-gravity of the field 
or the effect of the inflaton potential truncates the Hilbert space so that $N_{\cal H}$ is finite, which also implies a
    discreteness of final field amplitude states. A  
mode constructed explicitly with  a discrete  lattice of amplitude eigenstates shows that only a small fraction $\approx N_{\cal H}^{1/2}$ of amplitude eigenvalues occur in
typical  vacuum fluctuations. 
The quantum state  of the field is described as an entangled state of many spatial modes, defining a relationship between  entropy and the number of modes.  
The covariant  bound on the entropy of inflationary fluctuations translates to a bound on the number of modes contributing to the final eigenstates. If the density of modes in $\vec k$ space is not too high, the effects of discreteness may be observable in the statistics of background radiation anisotropy, galaxy clustering, and eventually, directly detected gravitational wave backgrounds.

\section{ Quantum Mechanics of Field Modes}

\subsection{Single Mode During Slow-Roll Inflation}

Adopt  the notation of ref.\cite{liddlelyth}.
In conformal variables, the  amplitude $u=a\delta \phi$ of a plane-wave mode of a massless field  during slow-roll inflation obeys the same classical equation of motion as a harmonic oscillator:
\begin{equation}
\ddot u + \omega (\tau)^2 u=0.
\end{equation}
Here $\delta \phi $ denotes the classical field pertubation amplitude, $a\propto e^Ht$ denotes the cosmic scale factor, $\dot{}$ denotes differentiation with respect to conformal time $\tau$ (where 
$d\tau=dt/a$, and $\tau= -(aH)^{-1}$), the time-varying oscillator  frequency is
\begin{equation}
\omega^2=k^2-2(aH)^2,
\end{equation}
where $k$ denotes comoving wavenumber (and $k/a$ the physical wavenumber), and $H$ denotes the expansion rate or Hubble parameter during inflation.

In the usual (Born-Heisenberg-Jordan) field quantization, the state of the mode is represented by an operator ${\bf u}$,
\begin{equation}
{\bf u}= w {\bf a} + w^* {\bf a}^\dag,
\end{equation}
where ${\bf a}$ and $ {\bf a}^\dag$ represent the usual particle annihilation and creation operators, and $w$ obeys the classical equation of motion, with an approximate solution (assuming $H$ constant),
\begin{equation}
w=(2k^3)^{-1/2}(i-k\tau)\tau^{-1}\exp (-ik\tau).
\end{equation}

The main statistical prediction of  inflation  is the spectrum of the perturbations. Since the rms field amplitude is a statistical classical observable, its expectation value  also obeys the classical equation of motion. The solution at late times for a field initially in the vacuum state is $|w|^2= a^2H^2(2k^3)^{-1}$. The  spectral density of fluctuations in the field $\phi$ then turns out to be $P_\phi= (H/2\pi)^2$, independent of the number of modes contributing--- and therefore independent of whether the spectrum is continuous or discrete.  The distribution of amplitudes in each mode is Gaussian (from the ground state wavefunction of a harmonic oscillator), also independent of the number of modes. 
 
To illuminate the process by which the quantum field fluctuations become frozen in classically, it is useful  instead  to follow Dirac-Schr\"odinger  quantization. The classical  Hamiltonian for the field mode is 
\begin{equation}
H_c= {1\over 2}\left({{\partial u}\over {\partial t}}\right)^2+ {1\over 2}{\omega^2\over a^2} u^2,
\end{equation}
leading to the Schr\"odinger  equation (in units with $\hbar=1$),
\begin{equation}
-i {{\partial \psi}\over {\partial t}}= -{1\over 2}{{\partial^2 \psi }\over {\partial u^2}}+ {1\over 2}
[ (k/a)^2-2H^2] u^2 \psi
\end{equation}
where $\psi(u)$ is the quantum-mechanical amplitude for the field mode to have the field amplitude $u$.

When the mode is much smaller than the apparent horizon radius $H^{-1}$, this is just the familiar quantum harmonic oscillator. The eigenstates correspond to definite oscillator energy, in this case to definite numbers of quanta $n$. The ground or vacuum state has zero quanta, but has the Gaussian $\psi(u)$ associated with zero-point fluctuations. 

As $k/a$ falls below $H$, the field quanta become more localized.  At  very late times, the Schr\"odinger  equation becomes
\begin{equation}
i {{\partial \psi}\over {\partial t}}\approx  H^2 u^2 \psi.
\end{equation}
The wavefunction at late times is a superposition of eigenstates of $u$,
\begin{equation}
\psi(u)=\sum_j  a_j \delta(u-u_j) \exp [- i H^2 u_j^2 t].
\end{equation}
 Eigenstates correspond to definite values of $u$.  A state which starts off in the vacuum eigenstate  at early times   ends up  as a superposition of  field-amplitude eigenstates at late times, with   Gaussian  coefficients corresponding  to the previously-derived rms value.
At late times we do not observe this superposition, but only one of the $u$ eigenvalues  $u_j$ with probability  $|a_j|^2$, with the standard Gaussian amplitude distribution. 
 Even though the quantum state of the  mode evolves coherently until long after inflation is over, its observable information content  is frozen very early (when $k/a\approx H$, a time we denote by $t_k$).

\subsection{Single mode with Pseudo-Self-Gravity}

In   standard inflationary field theory, the field is
embedded in a spacetime that acts like an apparatus; it has not been treated quantum-mechanically, but  is responsible for the change in $\omega$. A rudimentary
way to ask about something like quantized spacetime degrees of freedom is to add an operator to the Hamiltonian that represents the interaction and quantum entanglement of the spacetime and
field. We do this in two ways: first by adding an operator representing
 the self-gravitational energy 
of the field eigenstates, and then by adding  an operator corresponding to the energy of the inflaton potential.

Suppose we were to add another term to the oscillator Hamiltonian, corresponding to the self-gravitatational energy of the field. The order of magnitude of this term for a mode in one of the number eigenstates  would be
\begin{equation}
H_G\approx -G M^2/R\approx -{k\over a}{{E_n^2}\over m_{Pl}^2},
\end{equation}
where $E_n=(n+1/2) E_0$ denotes the oscillator energy in the $n$th number eigenstate, 
and $E_0$ denotes the zero-point energy.

Adding a negative term to the Hamiltonian produces a qualitative change in its spectrum:  the number of eigenstates spanning the Hilbert space ${\cal H}$ of  solutions (the
dimension $N_{\cal H}$) changes from infinite to finite.   For small $n$ and
$\omega<<m_{Pl}$, the gravitational self-energy term is negligible, and there are $\psi_n(u,t)$ eigenstates that closely resemble the number eigenstates of the simple harmonic
oscillator, with small modifications (such as small departures from Gaussianity).   However,  at higher
$n$,  $H_G$ comes to be significant. Since the gravitational energy term is negative, the harmonic-oscillator-type 
eigenstates are seriously altered above the point where ${ H}_G  \approx n E_0$.

For a mode that is freezing out (at $t_k$), this happens for
\begin{equation}
n_{G}\approx (m_{Pl}/H)^2.
 \end{equation}
Physically, at  this value, the standard eigenstate energy corresponds to the energy of a black hole of  size $\approx H^{-1}$;  an oscillator in that state has enough quanta to collapse to a black hole.  Although the model does not make
physical sense once it is pushed to this extreme, it nevertheless shows how gravity changes the Hilbert space of the field oscillator,  and gives an estimate
$N_{\cal H}\approx n_{G}$ of  the maximum number of eigenstates in the Hilbert space of a  quantum oscillator before quantum gravity becomes important.

\subsection{Single Mode with Inflaton Effective Potential}

The previous model    applies to any field that obeys quantum mechanics and couples to gravity. Now   consider an
effect that applies specifically to the  inflaton, because of its special role in controlling the dynamics of the spacetime.

It is standard in inflation theory to define an  effective potential $V(\phi)$, defined as the free energy density  corresponding to a  background vacuum expectation value
$\phi$ (in the  zero-momentum mode).  For wavelengths $<<H^{-1}$  we can discount this potential in considering spatial perturbations.  However, at $t>t_k$  the perturbation   leads to
definite frozen-in quasi-classical perturbations in $u$ larger than the horizon. These are frozen into the metric, with parameters related through the Friedman relation
$H^2=8\pi V/3 m_{Pl}^2$.  Self-consistency  suggests that as the mode makes the transition to zero momentum, we should include the perturbed inflaton effective potential energy,
as well as the gravitational self-energy, in the quantum-mechanical Hamiltonian.

A rough guess at this term around the time $t=t_k$ is
\begin{equation}
{ H}_V \approx H^{-3}V'u/a,
 \end{equation}
where $V'=\partial V/\partial \phi$, corresponding to the perturbation in free energy density integrated over a volume $H^{-3}$. Note that unlike self-gravity,  the sign of
this term depends on whether
$V'u$ is positive or negative. 

As before, consider the effect of this operator on the number eigenstates of the simple harmonic oscillator. Again, for small $n$ there is little effect, but at
large $n$ we find that ${H}_V  $ becomes comparable to $E_n $.  The series of  discrete  bound
state  solutions  is significantly modified at $n_V\approx (H^{-3}V')^2$. 

This  suggests a limit to $N_{\cal H}$  for the inflaton
perturbations, which in general can be more restrictive than the previous self-gravity limit. The combination of inflaton-potential
 parameters $H^{-3}V'$, in slow-roll inflation, is approximately given by the inverse of the final (observed) amplitude of gravitational metric scalar perturbations. Up to factors of the order of unity\cite{Hogan:2002xs}, $n_V\approx (H^{-3}V')^2\approx  10^{10}$. This gives an estimate $N_{\cal
H}$ for modes of the inflaton on the horizon scale at any given time--- approximately,
 the number of possible classical values   the frozen-in inflaton field amplitude can adopt during an
$e$-folding of inflation.

This agrees with an earlier argument\cite{Hogan:2002xs} from another point of view. Suppose that inflation proceeds in discrete steps, such that each state differs from the next by one
bit of entropy or one effective degree of freedom.  The bound on the total entropy of the inflationary causal diamond is $S_{max}=\pi
m_{Pl}^2/H^2\approx n_G$. In slow-roll inflation, in each $e$-folding of expansion, the area of the de Sitter horizon in Planck units increases by $\approx (H^{-3}V')^2\approx n_V$. This 
consistency is  reassuring since it suggests that the simple one-dimensional oscillator models are capturing  some of  the same essential physics as the general holographic arguments.

 \subsection{Single Mode with a Lattice of Amplitude States}

We have just seen how physical effects lead to a natural limit on $N_{\cal H}$ for field perturbations. We now examine some simple quantum-mechanical
systems with discreteness built in from the start, where we can  display explicitly the relationship between discrete spectra in $u$ and in $E$. This exercise is also useful for estimating how many of the system degrees of
freedom participate in the vacuum fluctuations relevant for inflation.

Consider a discrete system,   where $u$ adopts only discrete values $u_n$, separated by a uniform interval $b$, such that $u_n=nb$ where $n$ are integers.
Denote the wavefunction amplitude
to be at each value $u_n$  by
$\psi_n$, and the rate to  jump from one value to the next  $iA/\hbar$.
The Schr\"odinger equation of such an infinite 1-D lattice with no harmonic potential (that is, only the `momentum' part of the Hamiltonian) is\cite{feynman}:
\begin{equation}\label{eqn:lattice}
i\hbar{\partial\over \partial t} \psi_n= E_0\psi_n(t)-A\psi_{n+1}(t)-A\psi_{n-1}(t),
\end{equation}
where $E_0$ denotes the zero-point energy. Eigenstates  have the form
\begin{equation}
\psi(u_n,t)=e^{ik'u_n-(i/\hbar)Et}
\end{equation}
where
\begin{equation}
E=E_0-2A\cos k'b.
\end{equation}
The range of accessible energies goes from $E_0-2A$ to $E_0+2A$. For an infinite lattice, the range of internal wavenumber  $k'$ is continuous, but   a finite range ($2\pi/b$ in $k'$, which we may take
to be from 
$-\pi/b$ to
$+\pi/b$) suffices to span the full space of solutions (that is, for  values of $k'$ outside of that range, the actual $\psi_n$ solutions are dual to those of $k'$ within it.)
If the lattice is an infinite line of
$u_n$, the Hilbert space is infinite, allowing continuous choices of
$k'$ within that range.

Now suppose the lattice is  finite. This is easily done by putting it in a periodic box, identifying states $n$ and $n+N$. The dimension of the Hilbert space
is then
$N_{\cal H}=N$, the number of distinguishable $u_n$ states. We require $\psi_n=\psi_{n+N}$, and therefore $e^{ik'Nb}=1$, or 
\begin{equation}
k'Nb=2\pi\times\ {\rm integer},
\end{equation}
so that $k'$ also adopts one of $N$ discrete values,
\begin{equation}
k'_j=j {2\pi\over N b}, \ \ j= 1,N.
\end{equation}
The finite Hilbert space in this case can be projected onto $u$ or $k'$; in either case, any measurement will yield one of a discrete set of $N$ values.
The energy spectrum is also discrete; the energies at small $j$ go like $(2\pi/N)^2A^2j^2$.  

Based on these examples we can describe what happens if  discreteness is imposed on a harmonic oscillator. Suppose  the values of $u$ are constrained to
discrete, evenly spaced values separated by $b$. Consider a system   described by  a Schr\"odinger equation  of the form (\ref{eqn:lattice}) with a harmonic potential term
added,
$(\omega/a)^2u_n^2\psi_n$.  Suppose the dimension of the   Hilbert space is $N_{\cal H}$.
The highest energy eigenstate  has an energy of about
$N_{\cal H}\omega/2$, (corresponding to the maximum energy,  imposed by the discreteness, at $k'\approx \pi/b$), and a width in
$u$ of about $N_{\cal H}b$--- encompassing, necessarily, at most  $N_{\cal H}$ discrete values of $u_n$.  The wavefunction in the
ground state of the oscillator has    a width of about $\Delta u\approx
 N_{\cal H}^{1/2} b$, that is, the $N_{\cal H}^{1/2}$  eigenvalues of $u_n$ nearest zero. 
The number of discrete eigenvalues corresponding to the typical vacuum
fluctuation, analogous to inflationary quanta, is about $N_{\cal H}^{1/2}$, far less than the full Hilbert dimension $N_{\cal H}$.

\subsection{Many  Entangled Spatial Modes}

The number of eigenstates relevant to observations is not  $N_{\cal H}$ for a single plane-wave mode, but  for  the Hilbert space governing the quantum behavior of all the degrees of
freedom of   fields that freeze out as classical perturbations.  Although this number is much larger than for a single mode, it is still not infinite if
subject to gravitational or  holographic bounds. A similar simple model with  many entangled spatial modes approximates the  entropy implied by holographic bounds.

Instead of a single plane-wave mode, suppose the field on the Hubble scale is described as a combination of entangled spatial modes, where the quantum behavior of
each one   resembles  a simple harmonic oscillator (although in some examples below we will take them to be  smaller dimension quantum systems such as  binary qubits). Suppose that 
there are
${\cal N}$ such   modes, and that the quantum state of the whole field configuration is a sum over  eigenstates,
\begin{equation}\label{eqn:multimode}
 |\psi{>}=\sum_{i_1,i_2,...i_{\cal N}}a_{i_1,i_2,...i_{\cal N}}|\psi_1{>}\otimes |\psi_2{>} \otimes |\psi_3{>} \dots |\psi_{\cal N}{>}
 \equiv\sum_{i_1,i_2,...i_{\cal
N}}a_{i_1,i_2,...i_{\cal N}}|\psi_{i_1,i_2,...i_{\cal N}}{>}, 
\end{equation}
where $|\psi_i{>}$ label the states of each  mode. If $N_i$ denotes the number of states of each mode, $N_{\cal H}= N_i^{\cal N}$.
In standard field theory, the modes are  independent plane waves, and 
 both $N_i$ and ${\cal N}$ are infinite.

We have seen that for a single oscillator with self-gravity, $N_i$ is  limited to about $ m_{Pl}^2/H^2$, because states above that correspond to numbers of quanta that
would collapse to a black hole. A heuristic interpretation of holographic entropy bounds is that
  gravity entangles the multiple field modes.  The  system is again limited to states that involve no more
than about
$ m_{Pl}^2/H^2$ total quanta in a Hubble volume. Assuming the ${\cal N}$ modes are similar in wavelength (say, spanning an octave of $t_k$) and phased so that their gravity adds
coherently over a Hubble volume, the self-gravity operator on each eigenstate  contributes to the entangled Hamiltonian   a term of order
\begin{equation}\label{eqn:multigrav}
{  H}_G \approx (H/m_{Pl}^2) E_{i_1,i_2,...i_{\cal N}}^2 ,
\end{equation}
where $E_{i_1,i_2,...i_{\cal N}}\approx H {\cal N}$ for a  typical  state with about ${\cal N}$ modes of frequency $H$ excited.
The self-gravity of an individual eigenstate  becomes unphysically large (that is, forms a black hole on the Hubble
scale) unless  ${\cal N}< m_{Pl}^2/H^2$. That is, once the field is entangled with gravity, states with ${\cal N}> m_{Pl}^2/H^2$ are unphysical and should not be included in the Hilbert
space. All observables,  including the spatial structure of the
 field configurations, then  have   discrete spectra  with  a total of $N_{\cal H} = N_i^{\cal N} $  eigenstates. 
 Self-gravity  thus imposes a
limit  of about
$S=\log N_{\cal H}<
\log (N_i) m_{Pl}^2/H^2$. This  approximately agrees with the holographic entropy bound for the
inflationary event horizon,
$S<\pi m_{Pl}^2/H^2$. 

As in the case of a single mode, inflation changes the eigenstates of the multi-mode Hamiltonian. As modes pass through the horizon, the eigenstates change:
\begin{equation}
\sum_{i_1,i_2,...i_{\cal
N}}a_{i_1,i_2,...i_{\cal N}}|\psi_{i_1,i_2,...i_{\cal N}}{>}
\rightarrow \sum_{i_1,i_2,...i_{\cal
N}}a'_{i_1,i_2,...i_{\cal N}}|\psi'_{i_1,i_2,...i_{\cal N}}{>}. 
\end{equation}
Because  the evolution is unitary, the number of eigenstates does not change. 
Each  single mode changes to eigenstates  of amplitude, so the spatial structures of the  final eigenstates of the  many-mode system  is a superposition of quasi-classical plane waves.  
The   conjecture   that leads to the conclusion that spectral discreteness may be observable  in spatial structures  is that the final Hamiltonian eigenstates
$|\psi'_{i_1,i_2,...i_{\cal N}}{>}$  still have the separable form
\begin{equation}
|\psi'_{i_1,i_2,...i_{\cal N}}{>}=|\psi'_1{>}\otimes |\psi'_2{>} \otimes |\psi'_3{>} \dots |\psi'_{\cal N}{>}
\end{equation}
where the $ |\psi'_i{>}$ denote states of a final discrete set of  spatial modes.

\section{Entropy Bounds and Classical Outcomes}
Heuristic arguments aside,  covariant 
entropy bounds\cite{bousso,bousso02} lead  to a  general, quantitative upper limit on the entropy within the apparent horizon in slow-roll inflation. Define a
 light-sheet   volume 
$\cal V$ by taking a two-dimensional spacelike surface, say  a sphere $\cal S$  at some time
$t_0$ (defined in the standard FRW-like inflationary coordinate system),  and propagating it along inward-directed null trajectories towards the future and past 
The surface is chosen with a radius just slightly smaller than the horizon at radius
$H^{-1}$, such that the null surface defining $\cal V$ is everywhere converging. Then the covariant entropy bound\cite{bousso,bousso02} on the entropy of  $\cal V$ is:
\begin{equation}
S_{max}({\cal V})< A({\cal S})m_{Pl}^2/4\approx \pi  m_{Pl}^2/H^2.
\end{equation}

A conceptual illustration\cite{'tHooft:1999bw} of the holographic entropy  bound for black holes is to imagine the event horizon tessellated with pixels, each one of area $4\ln 2
m_{Pl}^{-2}= 0.724\times 10^{-65}{\rm cm^2}$. Each one of these represents an entangled qubit; any classical outcome must yield a 0 or 1 for each pixels. The number of squares is one
quarter of the area of the event horizon, and this is the entropy of the system $S$ in bits. The number of classical outcomes of an evaporating hole, and the number of eigenstates for
the whole entangled system,  including evaporation products, is
$N_{\cal H}=2^S$.

Similarly, the covariant entropy bound imposes a limit on the number of independent inflationary perturbation modes whose information is frozen in as classical perturbations\cite{hogan04}. The modes can be thought of as a scale-invariant lattice of states in $\vec k$ space, where the mean separation $\Delta k$ of modes in the three dimensions of $\vec k$ space is bounded by  
\begin{equation}
{\Delta  k\over k}> \left[ {4\over 3}{H^2I_H\over {f_H m_P^2  }}\right]^{1/3}.
\end{equation}
Here $I_H$ is the entropy per mode frozen by $t_k$, and $f_H<1$ denotes the fraction of total entropy carried by modes at $k/a<H$.  It is possible that      bound is large enough to produce observable effects, and may even be of the order of unity.

This  lattice     in   comoving spatial-wavenumber space resembles 
 'tHooft's black hole illustration. 
Setting  $I_H=\log 2$  corresponds to one qubit per mode, $N_i=2$ in eq. (\ref{eqn:multimode}). Any
eigenstate or classical outcome is   specified by one  bit per mode in $k$ space.  A
finite set of   plane waves is available to assemble any possible classical outcome. Their eigenvalues form a set of  possible spectral ``emission lines'', with no
fluctuation power at frequencies in between.

Even if  the final classical wavenumber spectrum is not actually discrete, the    modes comprising the   quasi-classical states after inflation are not independent plane waves, but include correlations. 
The   (discrete) mode  amplitudes are strongly correlated within the line spacing so they are not independent as in standard inflation. Since  the 
modes become entangled by gravity at time
$\approx t_k$, the degrees of freedom that are frozen in are not the same ones that comprise the approximate eigenstates of the Hamiltonian to begin with.  The subspace and structure of the final spatial modes  depends on the detailed structure and dynamics of the fundamental modes
giving rise to the holographic bound. 
Lacking a theory of quantum gravity to tell us how this works in detail, it makes sense to  ask whether these effects might have an observable influence on  actual data.

\section{Phenomenology of a Discrete Perturbation Spectrum}

The inflationary
fluctuations in inflatons or gravitons, when they become quasi-classical, imprint their spatial structure  onto the spacetime metric. The limited-information spectrum survives today in
  phenomena   that preserve the spatial structure  of the primordial perturbations. We offer very rough estimates of the degree of discreteness $\Delta k/k$ observable    in cosmic background anisotropy,
galaxy clustering, and primordial gravitational waves.

\subsection{Background Anisotropy}

If $\Delta k/k$ is of the order of unity, holographic discreteness or correlations might already be showing up in  the current WMAP anisotropy
 data\cite{Bennett:2003bz,Komatsu:2003fd,Spergel:2003cb}.   It has been noticed that the data displays  significant statistical anomalies in the low order
multipoles\cite{deOliveira-Costa:2003pu,Eriksen:2003db} not in agreement with  standard inflation.
  Suggested explanations of these effects have included 
 customized inflation 
 scenarios,   string-inspired modifications of relativity, and nontrivial topology\cite{Cline:2003ve,Feng:2003zu,Bastero-Gil:2003bv,Contaldi:2003zv,Tsujikawa:2003gh,Luminet:2003dx}.  
A holographic interpretation  fits qualitiatively with the current results, and makes new predictions.

The three effects that seem statistically significant--- a low quadrupole amplitude (about a 1 in 20 coincidence), an unsually planar octopole (1 in 20), and 
 alignment between the principal axes of the quadrupole and octopole (1 in 60)--- are known to be natural consequences  of a toroidal cosmic topology, as a result of
 a discrete spectrum of wavenumbers projected onto the symmetry axis\cite{deOliveira-Costa:2003pu,Luminet:2003dx}.  Nontrivial
topology  on this scale is ruled out by other tests\cite{Cornish:2003db}, but a holographically discrete spectrum   can produce similar statistical anomalies;
for example, a low quadrupole is not very unlikely because
the sample
(or  cosmic)  variance is larger for a discrete spectrum than for continuous Gaussian noise. The
discrete holographic mode structures could also   create correlations between
multipole components that would be independent in the Gaussian case.

The  fitted coefficients  $a_{\ell m}$ of  the WMAP  quadrupole and octopole components ($\ell = 2$ and $3$) display  significant features that can be
interpreted as spectral gaps and ``emission lines'' (see \cite{deOliveira-Costa:2003pu}, table III).  In the frame defined by their principal axes, 2 of the 5 quadrupole components and 4
of the 7 octopole components are nearly zero. Almost all the fluctuation power is coming from the other 3 components in each case. The contrast between the lines and gaps is
large; the amplitude ratios are more than a factor of ten, so the spectral gaps contain less than one percent of the fluctuation power. Although a lower power in two components
can be partially discounted for each $\ell$ (because two degrees of freedom were used to align the axes to minimize them), such an extreme high-contrast distribution is  not typical if
these components are drawn from  a random Gaussian distribution. Spectral gaps of power are of course commonplace in discrete spectra.

If approximately scale-invariant holographic discreteness is creating these large scale anomalies, it should  lead to similar anomalies   replicated on smaller scales.
There is evidence for nongaussian effects on  smaller scales in the WMAP data, both from a localized wavelet analysis\cite{Vielva:2003et}  and from correlations in the $a_{\ell m}$
phases\cite{Coles:2003dw}.  These effects seem likely to  be artifacts or foreground effects, but some of them might be
 an
imprint of  discreteness. 

The observed anomalies suggest $\Delta k/k$ of the order of unity. What level of discreteness $\Delta k/k$ is resolvable in principle? As $\ell$ increases,
the number of independent observables increases like $\ell^2$ so the 
number of resolvable modes increases like $(k/\Delta k)^3\propto \ell^{2}$, where roughly $k\propto \ell$. On the other hand, at high $\ell$ the movement of matter and radiation at recombination, associated with acoustic wave propagation, smears out the phase coherence over roughly $\Delta k/k\approx \ell/200$. Very roughly this suggests that the maximum resolution occurs at $\ell\approx 200^{3/5}\approx 24$ and allows resolution $\Delta k/k\approx 200^{-2/5}\approx 0.12$.
Even though the precision of  anisotropy statistics (such the variance of $C_\ell$'s) at $\ell\approx 20$ are limited by ``cosmic variance'' or sampling error, close study of $a_{\ell m}$'s may show signature patterns  of a discrete spectrum.

\subsection{Galaxy Clustering}
Large-volume galaxy surveys also provide a way of directly estimating scalar perturbation mode amplitude, spatial wavenumber and phase. Since these surveys sample a three-dimensional
volume, they provide complementary information to the anisotropy maps. 
Also, they preserve a more accurate record of the comoving spatial location of primordial fluctuations, since the dark matter does not participate directly in the acoustic plasma oscillations before recombination.

What is the limiting resolution of galaxy surveys? The primordial phase information is lost on the scale where perturbations have become nonlinear, around $0.01 H_0^{-1}$ where $H_0$ is the current Hubble parameter. The number of resolvable modes $(k/\Delta k)^3$ is roughly the number of such nonlinear regions in any given survey. 
The largest current
surveys, such as the 2DF survey\cite{Percival:2001hw} and Sloan Digital Sky Survey\cite{sdss}, currently reach $(k/\Delta k)$ of the order of a few, comparable with the low order CMB multipoles, but on a very different comoving scale.   A larger and  deeper survey, including the entire contiguous volume  of galaxies extending to the Hubble length, could resolve down to $\Delta k/k\approx 0.01$. 

\subsection{Gravitational Wave Backgrounds}

In the future,  primordial gravitational waves may be detected directly with an interferometer at high frequencies. Although the contemplated instruments have poor angular resolution for stochastic backgrounds, they have very high frequency resolution that may  resolve spacings between discrete lines in the primordial spectrum better than the other techniques can.

Currently planned ground 
experiments and space missions\cite{Hogan:2001jn,Hughes:2001ch} do not have the sensitivity to detect the primordial graviton background except in unusual inflation models with blue
spectra. However, a discrete wavenumber spectrum concentrates the same total power in a narrower set of frequency bands and make the background easier to detect.

The frequency resolution for an interferometer is roughly $\delta \omega\approx T^{-1}$ where $T$ is the observation time, so the fractional resolution in frequency $\omega$ is $\delta\omega/\omega\approx \omega T$. With no angular resolution, the number of modes contributing to the total power per octave of frequency is about $(k/\Delta k)^3$. If the spectrum is actually discrete (as opposed to simply having strong correlations on the discreteness scale), 
then the power in the emission lines, for  a given total fluctuation power, is increased.
 If the line frequency  is stable and coherent over an observation time (as for example would be the case if there were truly fundamental frequencies in the system, drifting on a Hubble timescale), the power enhancement factor in a single resolution element is then as large as
about $\omega T ( \Delta k/k)^3$.

Using this enhancement factor with estimates of  the  sensitivity of the Laser Interferometer Space Antenna (LISA) and the expected level of inflationary gravitational waves\cite{Hogan:2001jn}, it appears that LISA may detect the primordial, zero-tilt inflationary background in the most favorable band (around 3 mHz) if $\Delta k/k$ is of the order of unity. A LISA follow-on mission optimized at  higher frequencies reaches a level of $\Delta k/k\approx 0.2$ at about 0.06 Hz.
 In a  very  ambitious      gravitational wave interferometer   with the sensitivity to detect and  fit out signals from  all foreground  astrophysical binary sources  and to detect the primordial graviton
background from  inflation at about 1 Hz (such as a proposed ``Big
Bang Observer'' mission), the frequency resolution could be as high as 
$10^8$ after several years of observation,   resolve up to $10^7$ different discrete modes, and reach $\Delta k/k\approx 10^{-8/3}\approx 0.002$.  The
statistical properties of  a discrete background are so different from the standard stochastic-background prediction that this possibility should  be considered in
early stages of mission design.

\acknowledgements

I am very  grateful to J. Bardeen, R. Bousso,  D. Kaplan, A. Nelson, and  M. Strassler for useful discussions. This work was supported by NSF grant AST-0098557 at the University of
Washington.

{}
\end{document}